\documentclass[]{aa} 
\titlerunning{And XXXVI}
\authorrunning{J. D. Sakowska et al}
\usepackage[utf8]{inputenc}

\usepackage{natbib}
\bibpunct{(}{)}{;}{a}{}{,}
\usepackage{gensymb}

\usepackage{pdflscape}
\usepackage{longtable}

\usepackage{graphicx}
\usepackage{txfonts}
\usepackage[colorlinks=true, linkcolor=blue, citecolor=blue, urlcolor=blue]{hyperref}

\begin{document}

    \title{Andromeda XXXVI: discovery of a new ultra-faint dwarf galaxy towards M31}

   \author{Joanna D. Sakowska
          \inst{1}\fnmsep\thanks{Corresponding author: \url{jsakowska@iaa.csic.es}},
          David~Mart\'inez-Delgado\inst{2,3}\fnmsep\thanks{ARAID fellow},  Michelle L.~M. Collins\inst{4}, Matteo Monelli\inst{5,6,7}, Giuseppe Donatiello\inst{8}, Amandine Doliva-Dolinsky\inst{4}, Isabel M.~E. Santos-Santos\inst{9, 10}
          }
\institute{
$^{1}$ Instituto de Astrofísica de Andalucía (CSIC), Glorieta de la Astronom\'ia s/n, 18008 Granada, Spain\\
$^{2}$ Centro de Estudios de F\'isica del Cosmos de Arag\'on (CEFCA), Unidad Asociada al CSIC, Plaza San Juan 1, 44001 Teruel, Spain\\
$^{3}$ ARAID Foundation, Avda. de Ranillas, 1D, E-50018 Zaragoza, Spain\\
$^{4}$ Department of Physics, University of Surrey, Guildford GU2 7XH, UK \\
$^{5}$ INAF-Osservatorio Astronomico d'Abruzzo, via Mentore Maggini s.n.c., 64100 Teramo, Italy \\
$^{6}$ IAC- Instituto de Astrof\'isica de Canarias, Calle V\'ia Lactea s/n, E-38205 La Laguna, Tenerife, Spain\\
$^{7}$  Departmento de Astrof\'isica, Universidad de La Laguna, E-38206 La Laguna, Tenerife, Spain \\
$^{8}$UAI -- Unione Astrofili Italiani /P.I. Sezione Nazionale di Ricerca Profondo Cielo, 72024 Oria, Italy \\
$^{9}$ Institute for Computational Cosmology, Department of Physics, Durham University, South Road, Durham, DH1 3LE, UK \\
$^{10}$ Leibniz-Institut f\"ur Astrophysik Potsdam (AIP), An der Sternwarte 16, D-14482 Potsdam, Germany
}

    \date{Received 30th March 2026; accepted 26th May 2026}

    \abstract
  {}
   {We present deep imaging of Andromeda XXXVI (And XXXVI), a dwarf galaxy discovered through visual inspection of the Pan-Andromeda Archaeological Survey, using observations obtained with the OSIRIS$+@$GTC instrument.}
   {The colour-magnitude diagram of And XXXVI shows a well-defined red giant branch (RGB). However, constraining a distance is challenging because the tip of the RGB is sparsely populated and no horizontal branch stars are found. The RGB is nevertheless well matched by an old (12.5 Gyr), metal-poor ([Fe/H] = -- 2.5) isochrone shifted to the distance of Andromeda (776 kpc). With a projected distance of $\sim$119 kpc from M31, And XXXVI is therefore likely a satellite of Andromeda.}
   {With $M_{V} \sim -5.9 \pm 0.1$, half-light radius $r_{\rm h} \sim 64 ^{+30}_{-19}$ pc, and an ellipticity $\epsilon \sim 0.015^{+0.032}_{-0.012}$, And XXXVI is one of the faintest of ultra-faint M31 dwarfs discovered to date.}
   {The discovery of And XXXVI adds to the faint end of M31’s satellite luminosity function, suggesting the presence of an even larger population of very faint satellites. Deeper space-based imaging and/or spectroscopic observations are needed to better constrain its position within M31’s halo. Combined with a detailed star formation history, such data would help determine whether its old, metal-poor stellar population indicates early quenching, similar to the trends seen in Milky Way satellites, and whether And XXXVI could be considered a reionisation fossil.}

   \keywords{galaxies: evolution – galaxies: dwarfs - galaxies: halos - dark matter
               }

   \maketitle

%

\nolinenumbers

\section{Introduction}

\begin{figure*}
\begin{center}
\includegraphics[width=1\textwidth]
{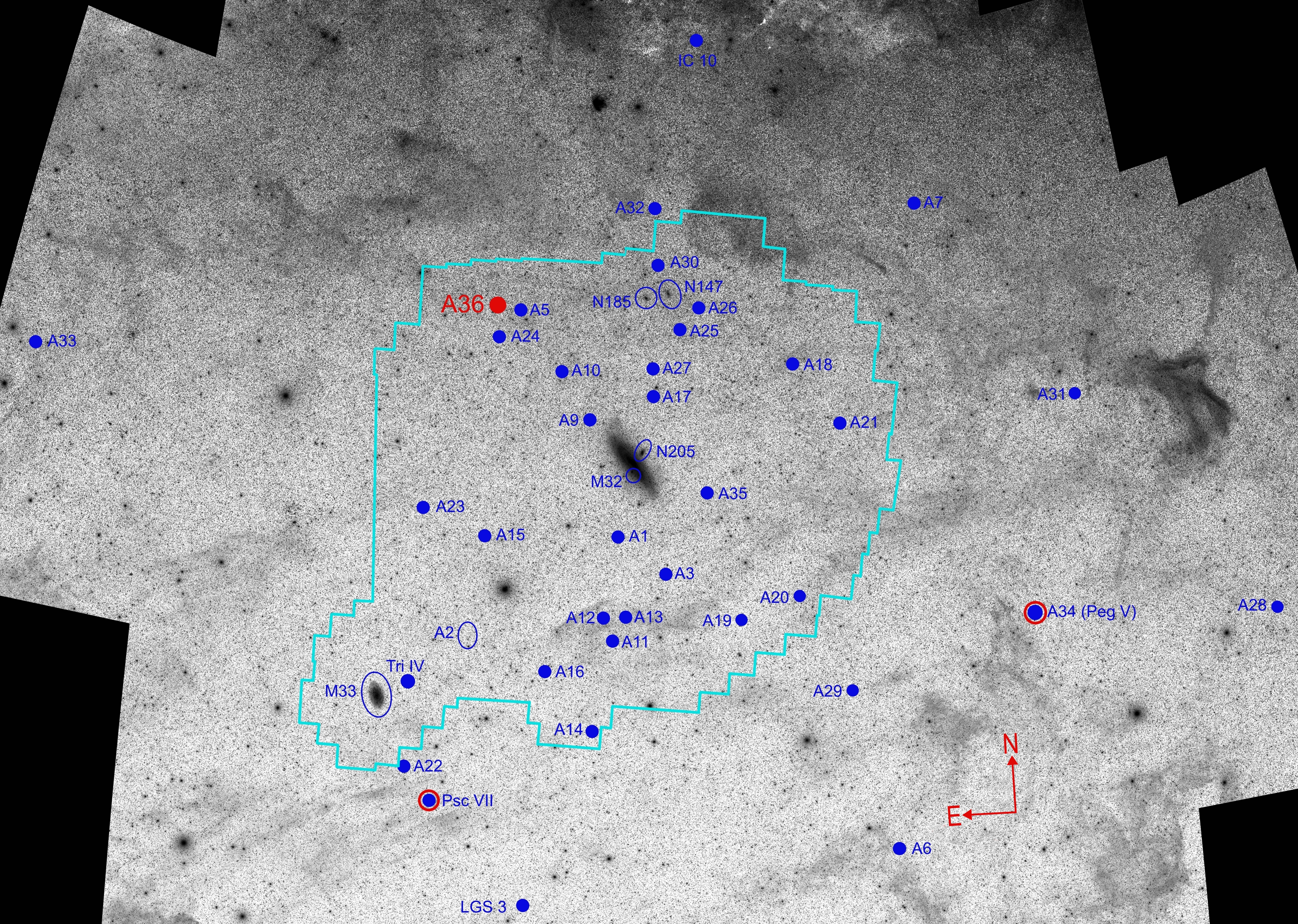}
\caption{Location of And XXXVI (marked in red) within the Pan-Andromeda Archaeological Survey (PAndAS; \citealt{McConnachie2018}), located approximately $\sim$ 119 kpc in projected distance from M31. We additionally mark Peg V and Psc VII, which were discovered following the same approach used in this work. The presented map is adapted from the And450 ultra-deep astrophotography survey (\citealt{Donatiello2025}).}
\label{fig:footprint}
\end{center}
\end{figure*}

\begin{figure*}
\begin{center}
\includegraphics[width=1\textwidth]
{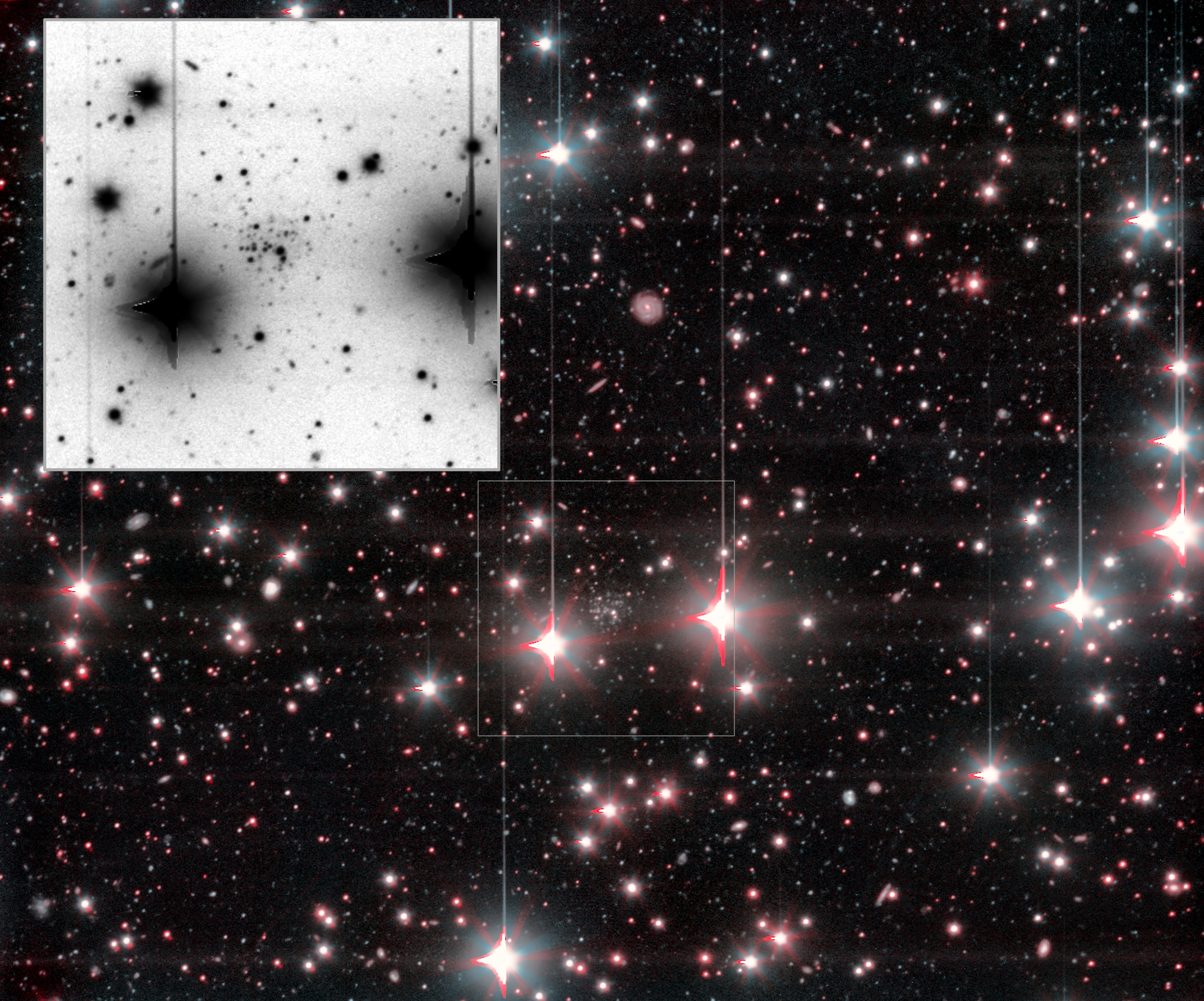}
\caption{Stacked OSIRIS+ image of And XXXVI with a field of view of 7.8 $\times$ 7.8 arcmin. In the inset we show a negative colour zoom-in on And XXXVI. The overdensity is clearly evident in-between two bright foreground stars. North is up, east is left.}
\label{fig:GTC}
\end{center}
\end{figure*}

Ultra-faint dwarf galaxies (UFDs; $M_{V}>$ --7.7, \citealt{Simon2019}) represent the extreme low-mass end of galaxy formation ($M_{*} \leq 10^{5}M_{\odot}$). Their shallow potential wells make them particularly sensitive to processes which regulate galactic growth, such as galactic outflows driven by stellar feedback (e.g., \citealt{Bovill2009, Applebaum2021, CollinsRead2022, rey2025}) or reionisation at early cosmic times (e.g., \citealt{Efstathiou1992,Bullock2000, brown2014, Weisz2014}). If these tiny galaxies formed the majority of their stars prior to the epoch of reionisation (and have undergone little to no evolution since) then they may be considered as reionisation fossils, offering precious constraints on pre-ionisation galaxy formation (e.g., \citealt{Gnedin2006}). Additionally, UFDs are some of the most dark matter-dominated systems known (e.g. \citealt{Bullock2017}), and their abundance and spatial distribution places observational constraints on the population of dark matter subhaloes around massive galaxies (e.g., \citealt{SimonGeha2007}). However, as a significant majority of these constraints have come from the Milky Way's (MW) own UFD population, studying external systems is thus important to avoid biasing our understanding of galaxy evolution and dark matter.

Given its proximity (776 kpc, \citealt{Savino2022}), and a massive dark matter halo ($M_{200}^{M31}\sim 2.0 \pm 0.5 \times10^{12} M_\odot$; \citealt{Fardal13}) expected to host plenty of substructures, the Andromeda galaxy (M31) provides an excellent test bed for identifying and characterising extremely low-luminosity systems. More than 40 dwarf galaxies are now known in the M31 system (\citealt{Pace2025}), of which $\sim$16 are classified as UFDs. This is a significant increase from the $\sim$10 M31 dwarfs found by the end of the century, which were mainly discovered through visual inspection of photographic plates (e.g, \citealt{VDB1972, Armandroff1998, Karachentsev1999}). The advent of wide-field photometric surveys, such as the Sloan Digital Sky Survey (SDSS; \citealt{SDSS}), the Panoramic Survey Telescope and Rapid Response System (PanSTARRS; \citealt{Chambers2016}), the DESI Legacy Imaging Survey (DESI LS; \citealt{Dey2019}) and UNIONS (\citealt{Gwyn2025}), has greatly accelerated the discovery of dwarf galaxies within and beyond the M31 halo ($\sim$150 kpc in projection). Thanks to their contiguous wide-field coverage, new dwarf galaxies were now discovered through both visual inspection of stacked images (e.g., \citealt{Slater2011}; \citealt{Martin2013a}; \citealt{Martin2013c}) and searches for resolved red giant branch (RGB) star overdensities (e.g., \citealt{Zucker2004, Martin2013b}) which were then confirmed as real systems through deeper photometric or spectroscopic follow-up (e.g., \citealt{Zucker2007, Bell2011, McQuinn2023}).

Yet the most substantial expansion of the known M31 satellite population has come from the Pan-Andromeda Archaeological Survey (PAndAS; e.g. \citealt{McConnachie2009, McConnachie2018}), carried out using the MegaCam wide-field imager on top of the Canada-France-Hawaii Telescope (CFHT). Contiguously mapping $\sim$150 kpc of M31 and $\sim$50 kpc of M33 to depths reaching three magnitudes below the tip of the RGB (TRGB), its depth and wide-field view accelerated galactic archaeology of M31’s halo, leading to the discovery of more than 20 dwarf satellites, down to luminosities of $M_V\sim-6$ (e.g, \citealt{martin2006, ibata2007, McConnachie2008, Martin2009, Richardson2011, Martin2016b}). Despite its depth, however, ground-based imaging of galaxies at the distance of M31 typically reaches only the red clump region of the colour-magnitude diagram (CMD). This makes it increasingly difficult to detect systems fainter than $M_V \sim -6$ as the number of RGB stars becomes considerably low, although very few discoveries have been recently made (\citealt{And35, Smith2025}). Identifying diffuse or partially-resolved overdensities with visual inspection, and following up candidates with deeper observations, therefore, remains a complementary approach to extending its observed satellite luminosity function (\citealt{Pace2025}, their figure 2). This approach is more sensitive to partially resolved (and unresolved) systems that could be otherwise missed by match-filter searches.

To make improvements in this direction, we have extended our visual search campaign to archival PAndAS data to search for semi-resolved candidates that may have been previously missed by automated searches. Our campaign has already identified UFDs in the outskirts of M31, such as Pisces VII (Psc VII; \citealt{MD2022, Collins2024}) and Pegasus V (Peg V; \citealt{Collins2022}), initially identified in shallower DESI LS data.  Their star formation histories (SFH) have since been found to suggest that both systems may have been quenched by cosmic reionisation (\citealt{Jones2026}), a scenario more commonly observed in MW satellites and potentially linked to differences in the accretion histories of the MW and M31 (e.g., \citealt{Sacchi2021, Savino2025}). These emerging differences between the MW and M31 satellite populations highlight the importance of expanding the census of UFDs across the Local Group.

In this paper, we present the discovery of Andromeda XXXVI (And XXXVI). In Section \ref{sec:meth} we describe our follow-up imaging with the OSIRIS+ instrument. Section \ref{sec:res} characterises And XXXVI, including its structural properties and luminosity. Finally, in Section \ref{sec:disc} we discuss And XXXVI and place our conclusions.

\begin{figure*}
\begin{center}
\includegraphics[width=1\textwidth]
{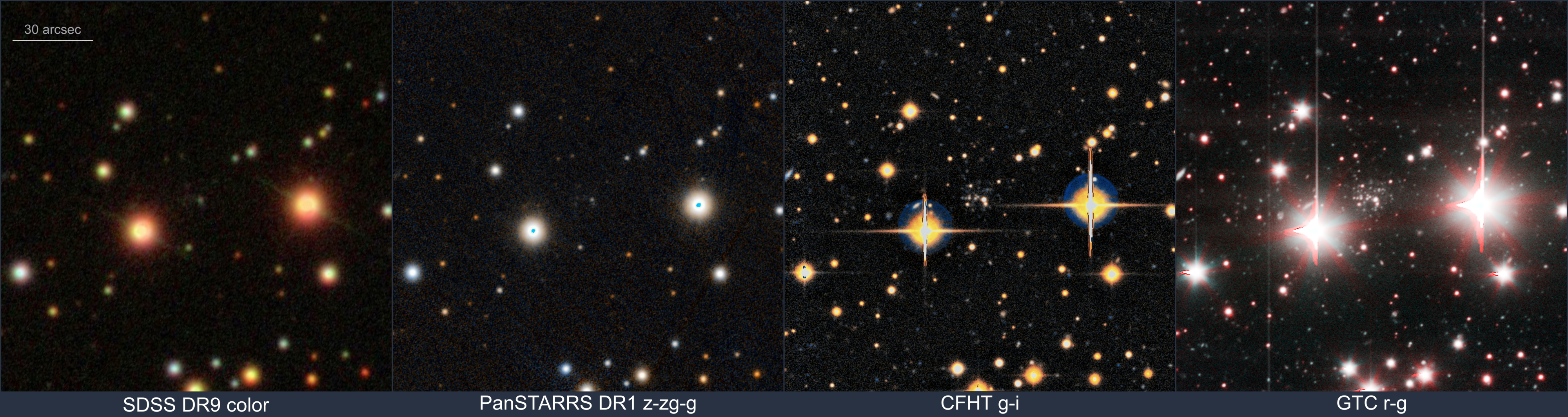}
\caption{Image of And XXXVI across four different datasets. From left to right: SDSS DR9, PanSTARRS DR1, CFHT (PAndAS) and GTC (this work). The overdensity is almost invisible in SDSS and PanSTARRS, and only becomes apparent with the CFHT and GTC.}
\label{fig:PAndAS}
\end{center}
\end{figure*}

\section{Observations and Data Reduction}
\label{sec:meth}


Andromeda XXXVI (And XXXVI) was discovered by amateur astronomer Giuseppe Donatiello during a systematic, visual inspection search of public images from the full PAndAS footprint (Figure \ref{fig:footprint}). The search followed the same approach as our previous studies with the DESI LS (M31: \citealt{MD2022, Collins2022}, Do I: \citealt{MD2018}, NGC 253: \citealt{MD2021, MD2025}). A total of 11 candidates were shortlisted, and we selected the two most conspicuous, semi-resolved overdensities for follow-up observations with the Director's Discretionary Time (GTC2025-292.227, PI: Mart\'inez-Delgado) using the 10.4-m Gran Telescopio Canarias (GTC; Roque de Los Muchachos Observatory, La Palma, Spain). In this work, we present our results for one of two candidates.

We took deep images of And XXXVI on January 22, 2026, with the Optical System for Imaging and low-Intermediate-Resolution Integrated Spectroscopy
(OSIRIS+)\footnote{\url{http://www.gtc.iac.es/instruments/osiris+/osiris+.php}} instrument on top of the GTC. We used an un-vignetted field of view (FOV) of 7.8'$\times$7.8' and a scale of 0.254" pixel$^{-1}$. The total exposure time was 20 $\times$ 150 sec = 3600 sec, and 20 $\times$ 149 sec = 2980 sec for the $g'$ and $r'$ photometric bands respectively. A 10-point dithering pattern (10$\arcsec$ offsets) was applied to correct for chip defects and improve image sampling. The seeing was around 0.9$\arcsec$ in $g'$ and $r'$. 

Images were processed using SAUSERO\footnote{\url{https://zenodo.org/records/15855515}}, a Python pipeline developed specifically to reduce OSIRIS+@GTC broad band imaging that is included in the standard GTC pipeline\footnote{\url{https://www.gtc.iac.es/observing/datareduction.php}}. Scientific images are pre-processed in the standard way (i.e. performing over-scan correction, bias subtraction, flat field division and astrometrisation via Gaia-DR3 stars), to produce a single final stacked image for each of the observed bands. The resulting image is shown in Figures \ref{fig:GTC}, \ref{fig:PAndAS}- the overdensity is now significantly more resolved than in the CFHT data belonging to the PAndAS survey. Prior to PAndAS, And XXXVI was almost invisible in PanSTARRS and SDSS due to its low-surface brightness.

\subsection{Photometry}
\label{sec:meth2}

\begin{figure*}
\begin{center}
\includegraphics[width=0.9\textwidth]{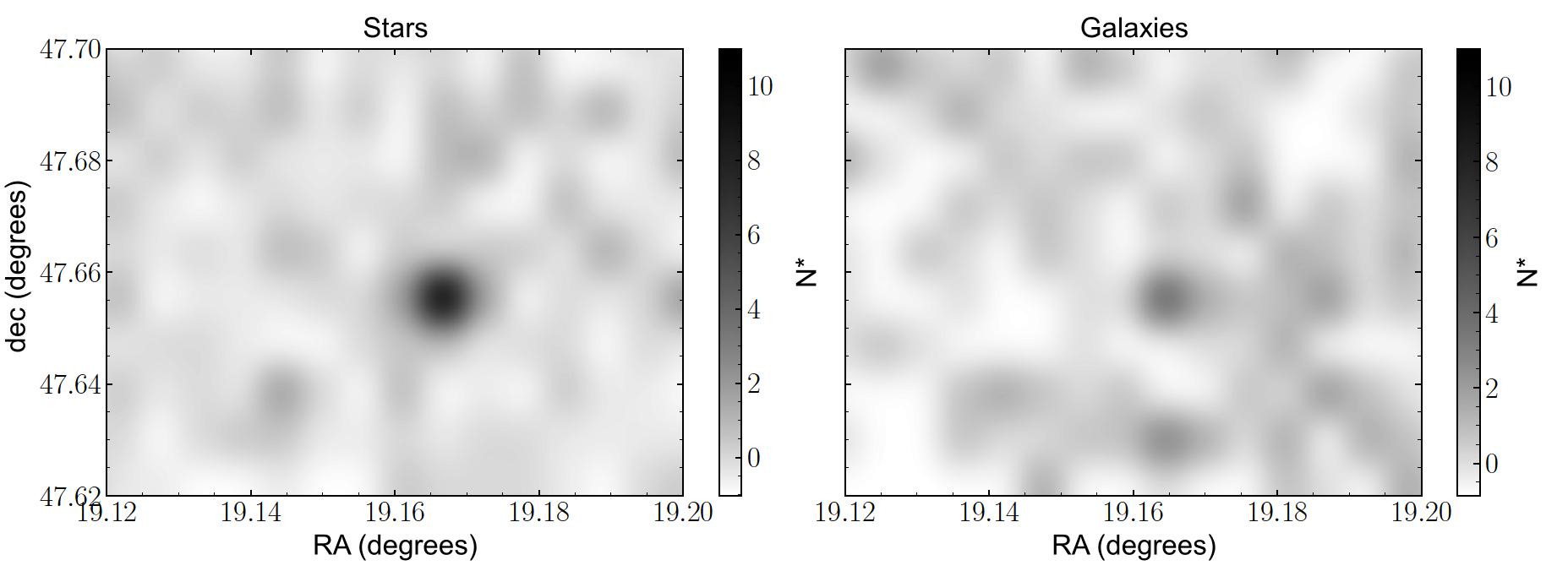}
\caption{Filtered maps of all sources in our GTC image categorised as `stars' (left) and `galaxies' (right) following our star/galaxy separation criteria. The overdensity clearly stands out in the stars panel.}
\label{fig:stargal}
\end{center}
\end{figure*}

\begin{figure}
\begin{center}
\includegraphics[width=0.5\textwidth]{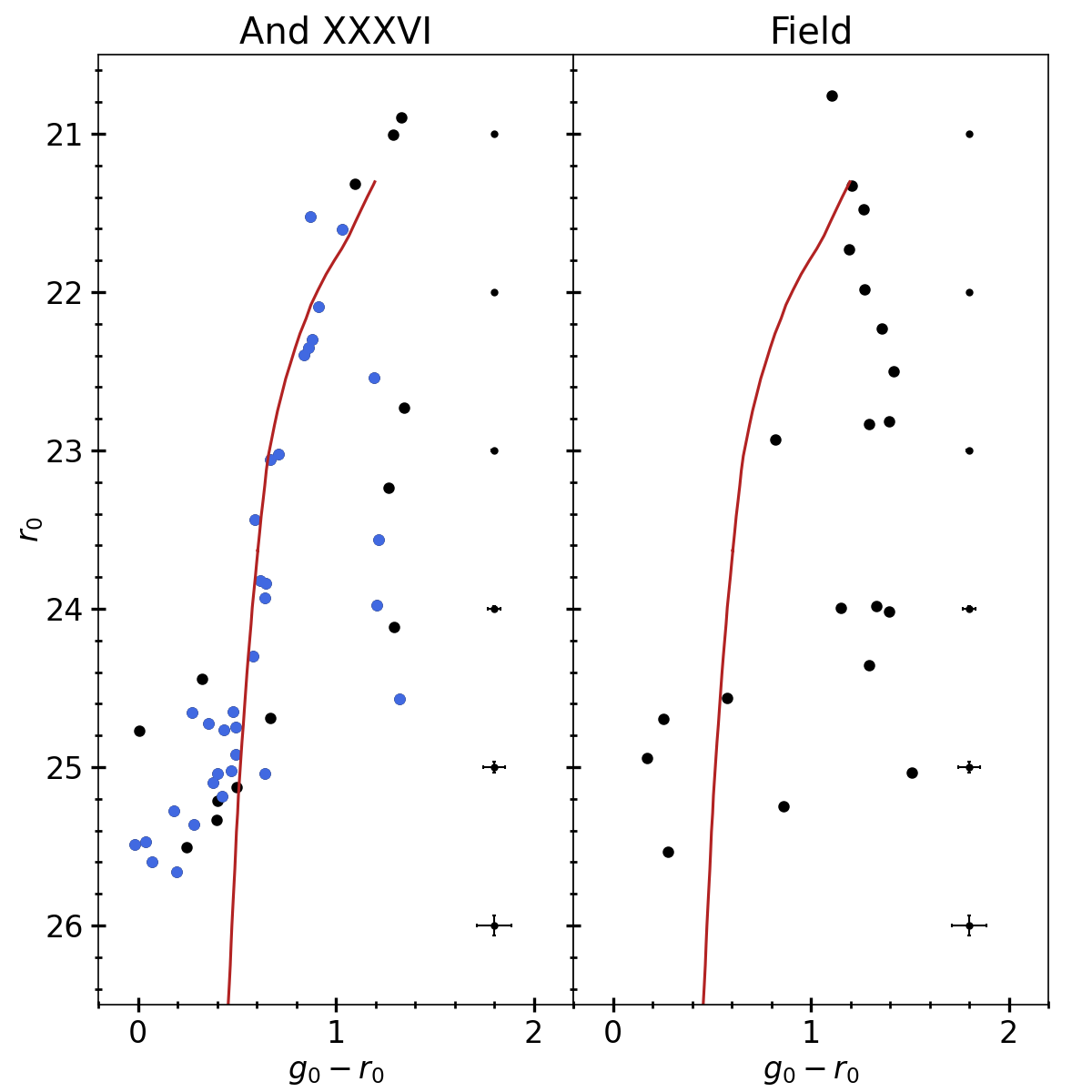}
\caption{Colour-magnitude diagram of And XXXVI (left) within 1$r_{h}$ (blue points, 0.284 arcmin radius) and 2$r_{h}$ (blue and black points, 0.568 arcmin). A clear RGB can be delineated. In red we overlay an old (12.5 Gyr), metal poor ([Fe/H] = $-$ 2.5) PARSEC isochrone using the distance to M31 (776 kpc). The isochrone matches the RGB locus well. On the right we show the CMD of an equal sized (2$r_{h}$) area near the overdensity, selected to represent our field contamination for visual reference (see text for details on field contamination determination)}.
\label{fig:CMD}
\end{center}
\end{figure}

The photometric reduction was performed using the DAOPHOT suite of codes \citep{stetson87}, performing the typical sequence DAOPHOT-ALLSTAR-ALFRAME. Aperture photometry in individual images was realised adopting a 3-$\sigma$ threshold. PSF stars were selected through automatic routines to pick up bright sources with low photometric error and good parameter shape, covering the full field of view to properly model the PSF spatial variations. After running ALLSTAR on individual images, ALLFRAME was used for the simultaneous reduction of all images, assuming an input list of stars obtained by merging the ALLSTAR catalogues. The PSF extraction and ALLFRAME steps were repeated twice more times, improving the fit quality as well as the geometric transformation between images. Special care was devoted to cleaning the input list from background unresolved sources and bad detections around saturated stars.

The photometric calibration curve was derived by a linear fit using stars in common with PanSTARRS. We assumed a relation in the form:
\begin{equation}
    \begin{split}
        g\_{cal} = g\_{instr} + 8.856 + 0.043*(g-r)\_{instr} \\
        i\_{cal} = i\_{instr} + 8.884 + 0.088*(g-r)\_{instr}
    \end{split}
\end{equation}

with r.m.s.$\sim$0.04 mag. We corrected the data for extinction using the \citealt{Schlegel1998} reddening maps, recalibrated by \citealt{Schlafly2011}. For our star/galaxy separation criteria, we selected stars that are measured by DAOPHOT to have $|0.2| <$ \verb|SHARP|. In Figure \ref{fig:stargal}, we recover the overdensity using this cut (left, 'stars' panel), showing it blends into the field when  $|0.2| >$ \verb|SHARP| (middle, 'galaxies' panel). 



\section{Characterising And XXXVI}
\label{sec:res}                                                          

\begin{figure*}
\begin{center}
\includegraphics[width=0.45\textwidth]{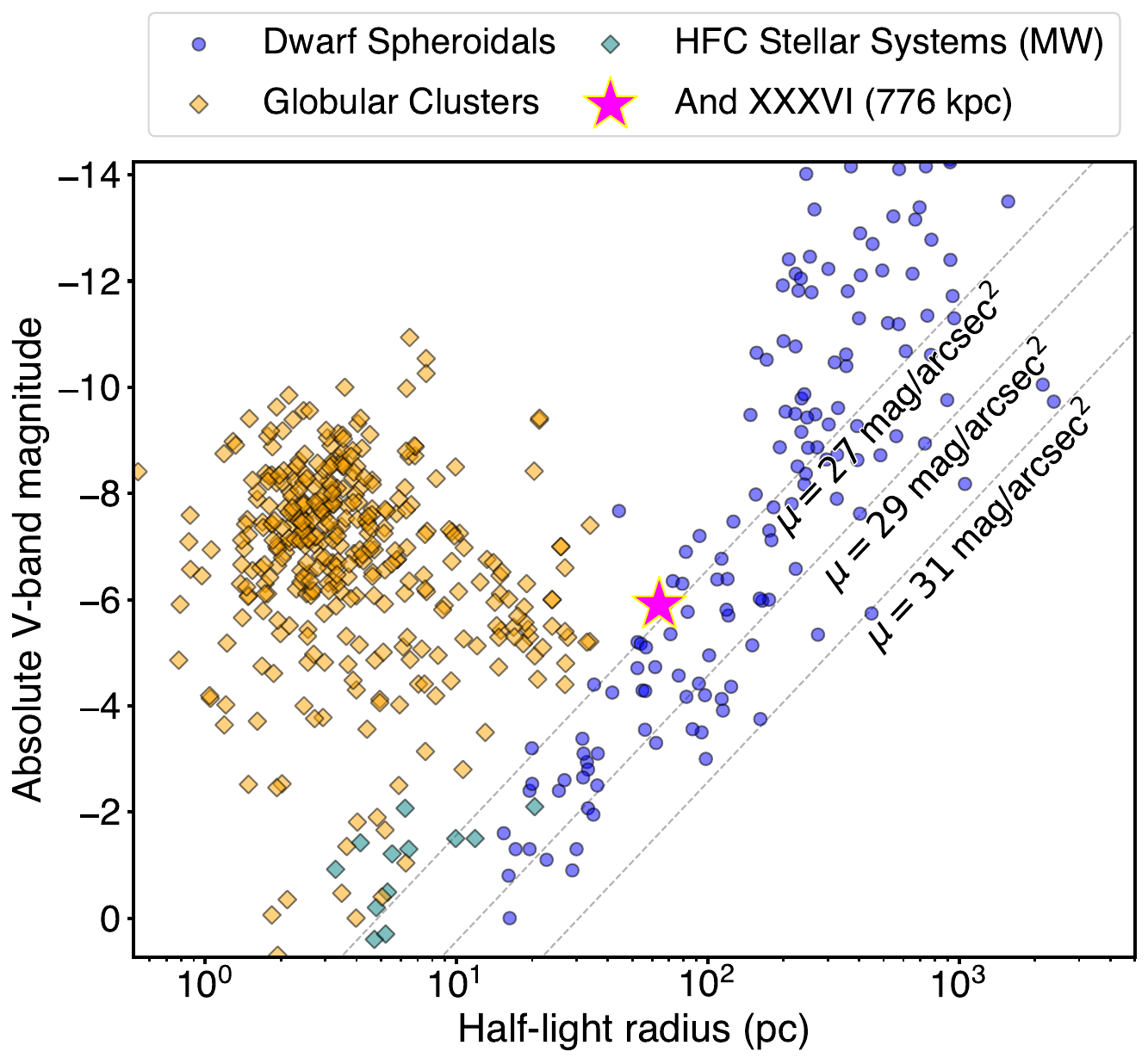}
\includegraphics[width=0.45\textwidth]{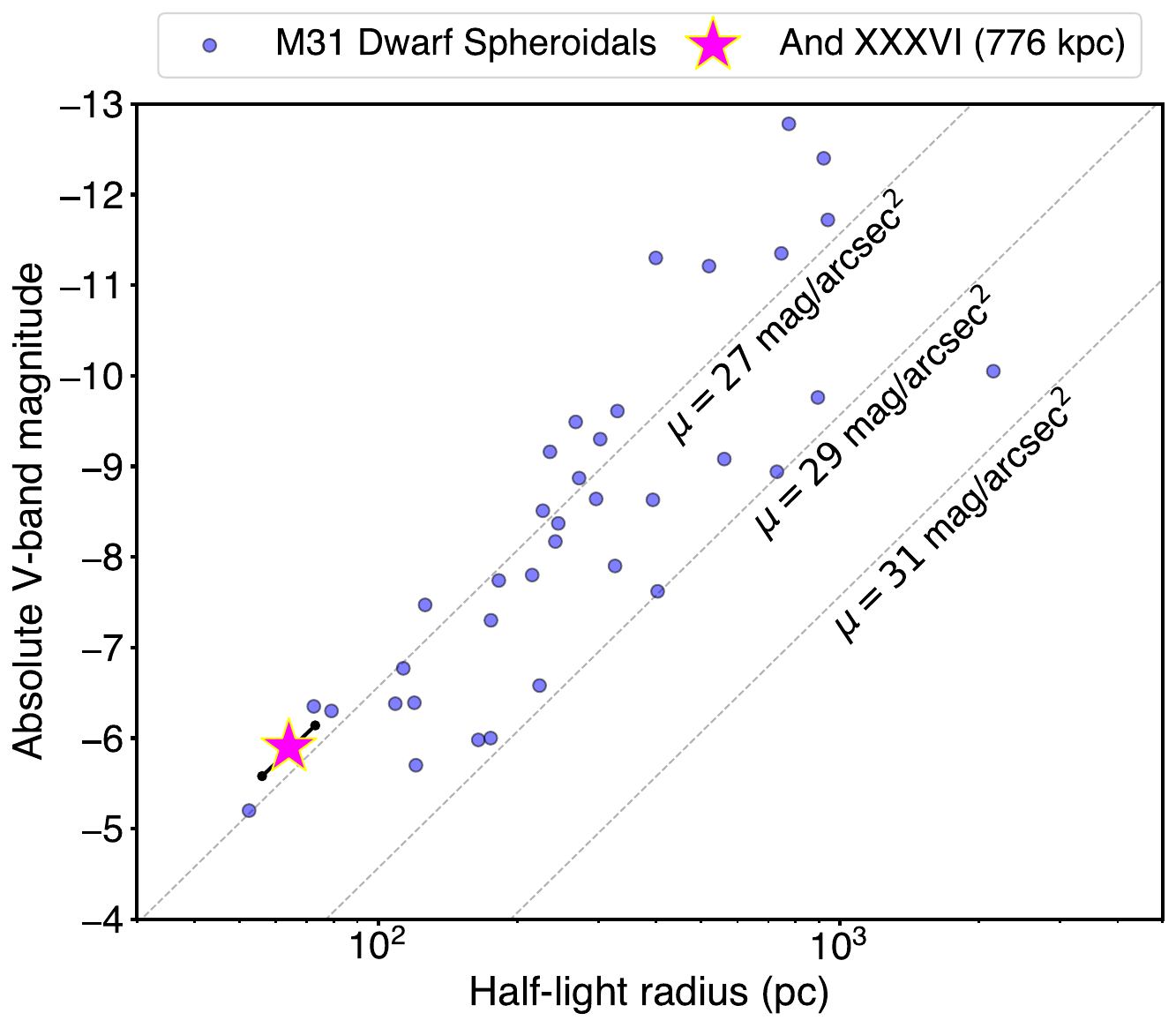}
\caption{Left: Absolute visual magnitude $M_{V}$ versus half-light radius $r_{h}$ (pc) for MW and M31 dwarf spheroidal galaxies, globular clusters, and MW only hyper-faint compact (HFC) stellar systems. We overlay our measurements for And XXXVI. Right: the position of And XXXVI versus other M31 only dwarf spheroidals on the same graph. In black we indicate the position of And XXXVI after varying the distance estimate by $\pm$100 kpc: And XXXVI fits securely on the size-luminosity relation as an ultra-faint dwarf galaxy. Data taken from the Local Volume database (\citealt{Pace2025}), selecting only systems confirmed to be real according to their criteria.}
\label{fig:Mv_rh}
\end{center}
\end{figure*}


In Figure \ref{fig:CMD} (left), we present the CMD of And XXXVI, selecting stars within a 0.284 arcmin (blue) and 0.568 arcmin (black) radius. To help disentangle And XXXVI from field stars, in the same Figure we show a CMD within an equally-sized (0.568 arcmin) area centred approximately $\sim$1 arcmin away (right). The RGB is clearly seen in the overdensity and not seen in the Field stars. As horizontal branch (HB) stars cannot be clearly identified, in red we overlay an old (12.5 Gyr), metal-poor ([Fe/H] = -2.5) isochrone from the PARSEC library (\citealt{Bressan2012}) at the average distance of M31 (776 kpc, \citealt{Savino2022}). While it is difficult to say whether the brightest stars ($r_{0} \leq 22$) are probable members or contaminating halo stars, the isochrone fits the RGB well overall.




To measure the luminosity and structural parameters of And XXXVI, we follow the methodology described in \cite{MD2022, Collins2022, Collins2024}. First, we apply the \verb|EMCEE| code (\citealt{Foreman-Mackey2013}) to determine the structural properties of And XXXVI, adapting the methods from \cite{Martin2016b}. Given EMCEE does not use distance or metallicity information, our structural parameter measurements are unaffected by different distance and metallicity assumptions. This approach assumes that the overdensity contains $N_{*}$ observed member stars with a radial density profile, $\rho_{\rm dwarf}(r)$, that can be described as:

\begin{equation}
    \rho_{\rm dwarf}(r)=\frac{1.68^2}{2\pi r_{\rm h}^2(1-\epsilon)}N^*\exp{\left(\frac{-1.68r}{r_{\rm h}}\right)}
    \label{eq:density profile}
\end{equation}

with $\epsilon$ representing its ellipticity ($\epsilon = 1- (b/a)$) and $r_{\rm h}$ is the half-light radius. The elliptical radius $r$ is defined as: 

\begin{equation}
    \begin{split}
        r= \Bigg( \Big(\frac{1}{1-\epsilon}((x-x_{0})\cos{\theta}-(y-y_{0})\sin{\theta})\Big)^2 \\
            +\Big((x-x_{0})\sin{\theta}-(y-y_{0})\cos{\theta}\Big)^2 \Bigg) ^{\frac{1}{2}}
    \label{eq:radial profile}
    \end{split}
\end{equation}

where $x_{0}$ and $y_{0}$ are coordinates for the candidate's centre, $x$ and $y$ are the coordinates on the plane tangent to the sky at the centre of the field, and $\theta$ is the position angle of the major axis. Throughout the analysis we describe the background contamination $\Sigma_{b}$ as constant. $\Sigma_{b}$ is determined by subtracting dwarf galaxy stars (calculated using equation \ref{eq:density profile}) from the total number of possible member stars:

\begin{equation}
    \Sigma_{b} = \frac{\left(n - \int_{A}^{} \rho_{\rm dwarf} \,dA\right)}{\int_{A}^{}dA} .
    \label{eq:background}
\end{equation}

Finally, we combine these equations \ref{eq:density profile}, \ref{eq:radial profile} and \ref{eq:background} to construct the likelihood function to be sampled by \verb|EMCEE|:

\begin{equation}
    \rho_{\rm model}(r) = \rho_{\rm dwarf}(r) + \Sigma_{b} .
    \label{eq:total profile}
\end{equation}

The \verb|EMCEE| sampler employs an iterative, Bayesian Markov chain Monte Carlo (MCMC) approach. We only feed in sources that fall within a colour range of $ -0.2 < g_{0} - r_{0} < 1.6$ and above our completeness magnitude cuts of $g_{0} < 25.5$, $r_{0} < 25$. To avoid over-constraining the solution we use broad, uniform flat priors- we set 0.5 arcmin for the half-light radius, $\pm1$ arcmin for the central RA and Dec, $0 < \theta < \pi$ for the ellipticity and $N_{*} \geq 0$. We also use 12 walkers, over a total of 30,000 iterations, with a burning stage of 20,000. In the Appendix Fig. \ref{fig:mcmc} we present the final corner plot provided by \verb|EMCEE|. And XXXVI is localised at RA = 19.168 and Dec = 47.656, with half-light radius $r_h = 0.284 ^{+0.132}_{-0.084}$ arcmin, ellipticity $\epsilon = 0.015^{+0.032}_{-0.012}$, position angle of the major axis $\theta = 78.6^{+29.7}_{-28.5}$ degrees and the number of stars $N_\mathrm{*} = 46^{+14}_{-11}$ corresponding to our CMD selection. The structural parameters obtained from \verb|EMCEE| are summarised in Table \ref{tab:mcmc_results}.

\renewcommand{\arraystretch}{1.4} 
\begin{table}[h]
\centering
\caption{Structural properties of And XXXVI derived from our analysis.}
\begin{tabular}{l c c}
\hline
Parameter & Value  \\
\hline
RA (deg) & $19.168 \pm 0.000$ \\
Dec (deg) & $47.656 \pm 0.001$  \\
$r_h$ (arcmin) & $0.284 ^{+0.132}_{-0.084}$  \\
$\epsilon$ & $0.015^{+0.032}_{-0.012}$\\
$\theta$ (deg) & $78.6^{+29.7}_{-28.5}$ &  \\
$N_\mathrm{*}$ & $46^{+14}_{-11}$  & \\
\hline
\end{tabular}\label{tab:mcmc_results}
\end{table}

\subsection{Luminosity of And XXXVI}

To calculate the luminosity of And XXXVI, we obtain a theoretical luminosity function from the PARSEC library representing a stellar population with an age of 12.5 Gyrs, metallicity [Fe/H] = $-$ 2.5 and $\alpha$ enhancement [$\alpha$/Fe]$ =+0.4$ dex, following our CMD result in Figure \ref{fig:CMD}. Using this information, we define the probability distribution function (PDF) which describes the expected RGB star count per magnitude bin. We model And XXXVI to have $N_*$ stars (taken from Table \ref{tab:mcmc_results}) and, assuming a distance of 776 kpc, we employ a probability-weighted random sampling of stars from the PDF above our completeness magnitude cuts. Once we have acquired $N_*$ stars, we convert all of the $g, r$ magnitudes to a luminosity and sum to get the final measurement. The average luminosity of And XXXVI was thus obtained by repeating this sampling process 1000 times. We recover $M_{g} \sim -5.4 \pm 0.1$ and $M_{r} \sim - 6.2 \pm 0.1$, respectively, calculating the error bar from the standard deviation of the individual results. By applying colour transformations of \citealt{Jordi2006}, we convert these values to absolute V-band magnitudes and obtain $M_{V} \sim -5.9 \pm 0.1$, which equates to a total luminosity of $L = (2.15^{+0.43}_{-0.36}) \times 10^{4} L_{\odot}$ for And XXXVI. The final luminosity and size parameters are summarised in Table \ref{tab:lum_results}, including the results obtained after shifting the distance assumption by $\pm$100 kpc.

\begin{table}[h]
\centering
\caption{Luminosity and size of And XXXVI according to the distance assumption.}
\begin{tabular}{l c c c c}
\hline
Parameter & 776 kpc (M31) & 676 kpc & 876 kpc  \\
\hline
$r_{\rm h}$ (pc) & $64 ^{+30}_{-19}$ & $56 ^{+26}_{-16}$ & $72 ^{+34}_{-21}$\\
$M_{g}$ & -5.4 $\pm 0.1$ & -5.1 $\pm 0.1$ & -5.6 $\pm 0.1$  \\
$M_{r}$ & -6.2 $\pm 0.1$   & -5.9 $\pm 0.1$   & -6.5 $\pm 0.1$\\
$M_{V}$ & -5.9 $\pm 0.1$ & -5.6 $\pm 0.1$ & -6.1 $\pm 0.1$ \\
$L(L_{\odot}) \times 10^{4}$ & $1.90^{+0.12}_{-0.13}$ & $1.44^{+0.09}_{-0.10}$   & $2.42^{+0.15}_{-0.17}$ \\
\hline
\end{tabular}\label{tab:lum_results}
\end{table}

\section{Discussion and conclusions}
\label{sec:disc}

We report the discovery of a new, semi-resolved object near M31 (And XXXVI), discovered using archival PAndAS CFHT data. Our follow-up GTC data has vastly improved the CFHT image of And XXXVI (Fig. \ref{fig:PAndAS}), resolving enough stars to construct its CMD and characterise its structural and luminosity properties. While we do not identify probable HB stars that would help anchor its distance, by overlapping an old (12.5 Gyr), metal-poor ([Fe/H] = $-$ 2.5) isochrone, we demonstrate that the RGB locus is consistent with the distance to M31 (776 kpc). Given its projected distance of $\sim$119 kpc from M31, And XXXVI would be located well within the estimated virial radius of M31 ($r_{200}\sim 260$ kpc, assuming an NFW profile of average concentration), suggesting it could be a bound satellite of this galaxy.

Assuming a distance of 776 kpc, we measure $M_{V}\sim- 5.9\pm0.1$ ($L = 1.90^{+0.12}_{-0.13} \times 10^{4} L_{\odot}$), and show that And XXXVI is relatively small with $r_{\rm h} \sim 64 ^{+30}_{-19}$ pc and $\epsilon \sim 0.015^{+0.032}_{-0.012}$. By comparing its derived size and luminosity to other confirmed compact systems around the MW and M31 (Fig. \ref{fig:Mv_rh}), we conclude that And XXXVI is most likely one of the faintest M31 dwarfs discovered to date. After varying our distance assumption by $\pm 100$ kpc (Table \ref{tab:lum_results}), we demonstrate that And XXXVI lies securely on the size-luminosity relation as an UFD. 
Follow-up kinematic data is required to decisively constrain And XXXVI's member stars and structural parameters. Targetting the brightest stars close to the TRGB would especially help constrain its distance (e.g. \citealt{Collins2024}) and confirm its $M_{V}$ measurement. If And XXXVI is confirmed to be old and metal-poor, it would be interesting to constrain the quenching time through a detailed SFH. If truly a reionisation fossil, it would question whether the SFH differences between M31's and the MW's satellite population are driven by environment. This would require deep space-based photometry (such as from the Hubble Space Telescope) to obtain a CMD reaching down to the oldest main-sequence turn-off (oMSTO, e.g., \citealt{Savino2023, Jones2026}) to understand when and how this system was quenched.

M31 is predicted to host as many as $\sim$92 dwarf galaxies ($M_{V} < -5.5$, \citealt{Doliva-Dolinsky2023}) yet $\sim$43 are currently known, suggesting that our census of M31 satellites is far from complete. Given that upcoming wide-field telescopes, such as the Legacy Survey of Space and Time (LSST; \citealt{LSST}) and Euclid (\citealt{mellier2025}), will not observe M31, follow-up observations of archival data with large ground-based telescopes (8+ meter class, under excellent seeing conditions) or space telescopes remains a promising avenue towards improving our knowledge of the M31 system. Indeed, the discovery of And XXXVI highlights that visual inspection remains very complementary to automatic and machine learning approaches, using resolved and/or semi-resolved data. Both methods in combination thus remain crucial towards constructing a complete picture of Andromeda.

\begin{acknowledgements} We thank the anonymous referee for helping improve the manuscript, and Antonio L. Cabrera-Lavers for his support during the DTT GTC observations. Data is available upon request to the corresponding author. JS and DMD acknowledge financial support from project PID2022-138896NB-C53. JS acknowledges financial support from the Severo Ochoa grant CEX2021-001131-S funded by MCIN/AEI/ 10.13039/501100011033. DMD thanks financial support for a visiting researcher stay at the Astronomy and Astrophysics Department of the University of Valencia within the framework of the «Talent Attraction» programme implemented by the Office of the Vice-Principal for Research (INV25-01-15). MLMC acknowledges support from STFC grants ST/Y002857/1 and ST/Y002865/1. MM acknowledges support from the Agencia Estatal de Investigaci\'on del Ministerio de Ciencia e Innovaci\'on (AEI-MCINN) under grants "At the forefront of Galactic Archaeology: evolution of the luminous and dark matter components of the MW and LG dwarf galaxies in the {\it Gaia} era" with references PID2020-118778GB-I00/10.13039/501100011033 and PID2023-150319NB-C21/10.13039/501100011033. ADD acknowledges support from STFC grants ST/Y002857/1.

Based on observations made with the GTC telescope, in the Spanish Observatorio del Roque de los Muchachos of the Instituto de Astrofísica de Canarias, under Director’s Discretionary Time GTC2025-292.
\end{acknowledgements}

\appendix

\begin{figure*}
\begin{center}
\includegraphics[width=1\textwidth]{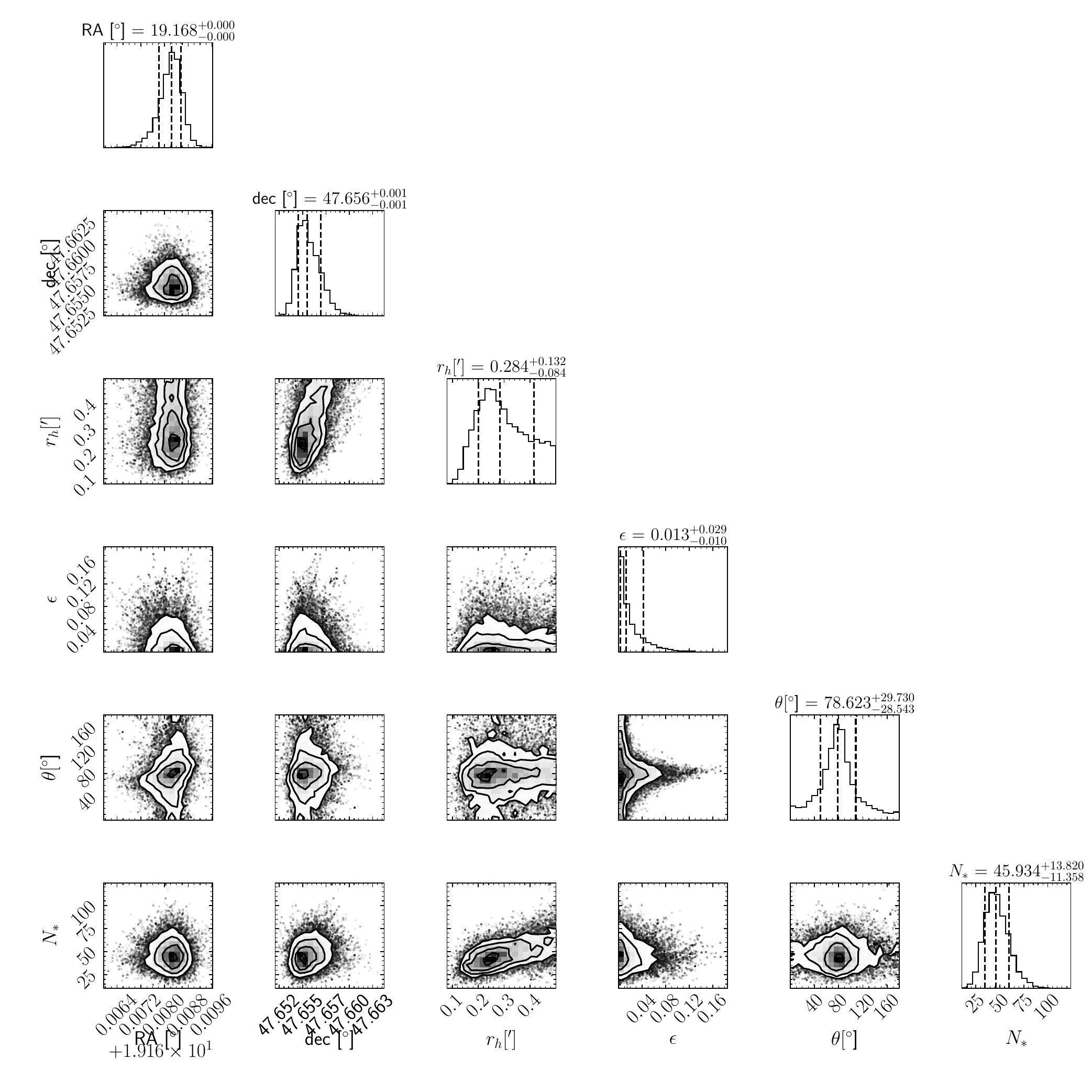}
\caption{Results of the structural analysis performed with EMCEE on And XXXVI- the central coordinates (RA, dec), half-light radius $r_{h}$ (arcmin), ellipticity $\epsilon$, the position angle of the major axis $\theta$ (degrees), and the number of stars $N_\mathrm{*}$ based on the CMD selection cuts. We use dashed lines to indicate the average result and 1$\sigma$ error boundaries.}
\label{fig:mcmc}
\end{center}
\end{figure*}

\bibliographystyle{aa} 
\bibliography{Bibliography}

\end{document}